# Loss of superconductivity and structural transition in $Mg_{1-x}Al_xB_2$

J.S. Slusky, N. Rogado, K.A. Regan, M.A. Hayward, P. Khalifah, T. He,
K. Inumaru, S. Loureiro, M.K. Haas, H. W. Zandbergen and R.J. Cava
*Department of Chemistry and Princeton Materials Institute*
*Princeton University, Princeton NJ 08544*

(February 14, 2001)

The basic magnetic and electronic properties of most binary compounds have been well known for decades. Therefore the recent announcement of superconductivity at 39 K in the simple binary ceramic compound $MgB_2$ [1] is surprising. This compound, available from common chemical suppliers, and used as a starting material for chemical metathesis reactions [2], has been known and structurally characterized since the mid 1950's [3]. Here we show that the addition of electrons to $MgB_2$ through partial substitution of Al for Mg results in the loss of superconductivity. Associated with the Al substitution is a subtle but distinct structural transition, reflected in the partial collapse of the spacing between boron layers near 10% Al content. This indicates that superconducting $MgB_2$ is poised very near a structural instability at slightly higher electron concentrations.

Electronic structure calculations on $MgB_2$ have shown that the states at the Fermi energy ($E_F$) are primarily derived from boron orbitals [4-6]. Doping on the Mg site is therefore expected to introduce electrons into the bands at $E_F$ with relatively little disruption of the electronic network. Solid solutions of the type $Mg_{1-x}Al_xB_2$ were therefore synthesized by direct reaction of the elements to test the effect of electron concentration on $T_c$. $AlB_2$ is the prototype compound for this structure type, and the $Mg_{1-x}Al_xB_2$ series has been previously reported [7] but not physically characterized. Starting materials were bright Mg flakes (Aldrich chemical), fine Al powder (Alfa Inorganics), and sub-micron amorphous B powder (Callery Chemical). Due to the poor reactivity of crystalline boron at low temperatures, and MgO contamination of fine Mg powders, the selection of starting materials is important. Starting materials were lightly mixed in half-gram batches, and pressed into pellets. The pellets were placed on Ta foil, which was in turn placed on $Al_2O_3$ boats, and fired in a tube furnace under a mixed gas of 95% Ar 5% $H_2$. The samples were heated for one hour at 600 C, one hour at 800 C, and one hour at 900 C. After cooling, they were pressed into pellets and fired for an additional 2 hours at 900 C and quickly cooled to room temperature. The resulting pellets have very low densities, making them inappropriate for resistivity measurements. Identical results to those reported here were obtained for materials synthesized by heating in sealed evacuated Ta tubes at 925 C for 4 hours.

The materials prepared by the above method in the composition region of interest for study of the superconducting properties, $Mg_{1-x}Al_xB_2$ for $0 \leq x \leq 0.40$, were all found to be solely of the $AlB_2$ type by powder X-ray diffraction. This crystal structure consists of honeycomb-net planes of boron, separated by triangular planes of the metals. $MgB_4$, the product expected for decomposition of $MgB_2$ due to Mg volatility at elevated temperatures [8], was not present. The materials were single-phase $AlB_2$ type in the lower and the higher concentration ranges studied. In the middle region, however, two-phase mixtures of two different $AlB_2$ type phases were observed, distinguished by significantly different $c$ lattice parameters. Partial representation of the powder diffraction patterns showing the (002) and (110) diffraction lines, characteristic of the inter-plane boron sheet separation and the in-plane boron net size, are shown in figure 1. The (002) reflection is sharp at both low and high Al concentrations, but becomes two broad peaks at intermediate concentrations. The ratio of those peak intensities changes with Al concentration in a manner characteristic of the mixing of two phases. Therefore the materials are single phase for x up to approximately 0.1, and again beyond x=0.25. The incompatibility of the two phases is strong enough in character to result in a distinct two-phase region between them. Elemental analysis by EDX in a transmission electron microscope (Philips CM200) indicated that the particles contained a bimodal distribution of Al concentrations at intermediate Al contents, even within very small particles (0.1 micron in some cases). The presence of both Al rich and Mg rich phases in the composition region near x=0.2 gives rise to the double-peak (002) structure and its dependence on concentration. A distribution of Al contents within those phases gives rise to the generally broadened (002) reflections observed in this region. The relatively smaller shift of the (110) reflection indicates that the in-plane size of the boron net does not change as dramatically as the inter-plane spacing at this structural instability.

The composition dependence of the crystallographic cell parameters was obtained by least-squares fitting to the positions of 9 powder diffraction peaks between 20 and 90 degrees 2Θ. The results are presented in figure 2.

The in-plane lattice parameter ($a$), which characterizes the dimension of the boron honeycomb net is relatively insensitive to Al concentration. Within the large family of metal diborides [9], the in-plane lattice parameters are relatively constant, reflecting the rigidity of the boron honeycomb layer. The separation between boron planes (the $c$ axis), on the other hand, changes significantly with Al substitution. There is a discontinuity in the $c$ axis length on passing through the two-phase region centered near x=0.2. In addition, $c$ varies differently with aluminum concentration within the two $AlB_2$ type phases (e.g. for $0 \leq x \leq 0.10$ and $x \geq 0.25$). The change in $c$ with Al concentration is significantly greater in the high Al content phase. This suggests the presence of a difference in electron orbital filling at low and high Al concentrations which is strongly coupled to the spacing between the boron layers.

The temperature dependent magnetization at low temperatures was measured for these materials, in the form of loose powders, in a Quantum Design PPMS magnetometer under an applied DC field of 15 Oe. The data, taken on heating after cooling in the absence of a field, are presented in figure 3. $MgB_2$ synthesized by the method described here has a very sharp superconducting transition, with a $T_c$ of 38 K. The transition is apparently one degree lower than that reported for samples synthesized with isotopically pure $B^{11}$, but is similarly narrow in width, and much sharper than that found for commercial powder [10]. Electron microscopy examination indicated that the particle sizes for the $Mg_{1-x}Al_xB_2$ powders were very small, on the order of a few tenths of a micron, and further that the two phases present for compositions in the two-phase region were intergrown on a nanometer length scale. Therefore the granular nature of the superconducting phase and possible proximity effects are expected to influence the shapes of the curves in figure 3, as is the presence of a distribution of Al contents in the powders. $T_c$ drops by a few degrees with electron/Al doping for Al contents less than 0.10 (from 38 K at x=0 to 36 K at x=0.1), and the transitions remain sharp. The transitions broaden significantly for Al concentrations greater than x = 0.10. The X-ray and magnetic data suggest that x=0.1 is barely within the two-phase region. The onsets of the transitions are all in the vicinity of 36 K, the transition temperature for x=0.10. There is no bulk superconducting transition for x=0.3 and x=0.4. The data are interpreted as indicating that bulk superconductivity in single-phase material is suppressed for x greater than 0.10. The superconducting transitions observed in the two-phase region are due to both the decreasing amount of the bulk superconducting phase with increasing Al content, and also to the distribution of Al contents in the powders. The structural and magnetic data taken together therefore indicate that the disappearance of bulk superconductivity in $Mg_{1-x}Al_xB_2$ occurs at the same Al concentrations at which a structural transition occurs which results in the partial collapse of the separation between boron planes.

The similarity of the calculated electronic density of states (DOS) for $MgB_2$ and $AlB_2$ [6] indicates that the effect of substituting Al for Mg can be considered primarily as a simple filling of available electronic states, with one electron donated per Al, within a "rigid band" picture. The calculations show [4-6] that there is a sharp drop in the DOS of $MgB_2$ at only slightly higher electron concentrations. The gradual decrease of $T_c$ from 38 to 36 K with increasing Al content in the single-phase region below x = 0.1 may therefore be due to the expected decrease of the DOS at $E_F$ with increasing electron count, consistent with a conventional origin for the superconductivity [10].

The substitution of Al for Mg in the high temperature binary ceramic superconductor $MgB_2$ decreases the superconducting transition temperature and leads to the loss of superconductivity. A structural transition is found at a composition corresponding to the disappearance of bulk superconductivity. High-resolution structural study of uniform powders, if they prove possible to synthesize, would be of interest to determine how the superconductivity and the structural transition are linked on an electronic level. The proximity of $MgB_2$ to a structural instability is consistent with a general picture for conventional high temperature superconductors as being near chemical phase instability due to strong electron-lattice coupling. However, our data for $Mg_{1-x}Al_xB_2$ do not suggest that the structural instability in this case is driven by competition between superconductivity and a structural distortion that decreases the density of electronic states at the Fermi Energy, as the Al substitution both decreases the DOS at $E_F$ and $T_c$. It is at first sight surprising that the observed structural instability in $MgB_2$ involves a partial collapse of the separation of the boron layers, not a change in the boron-boron in-plane distance. It will be of interest to determine whether this collapse is associated with special characteristics of the electronic band filling near 0.1 excess electron per cell.

## Acknowledgement
This work was supported by the U.S. Department of Energy, grant DE-FG02-98-ER45706, and partly by NSF grants DMR 972-5979 and DMR 980-8941.

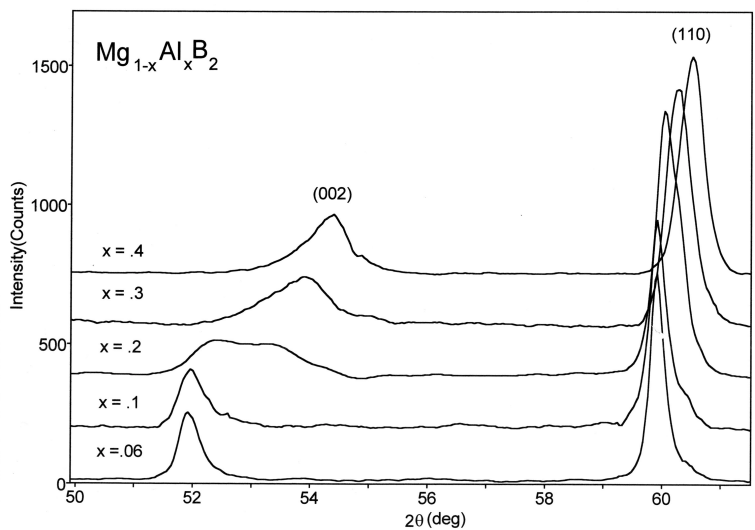

**Fig. 1**. Powder X-ray diffraction patterns (Cu K$\alpha$ radiation) in the region of the (002) and (110) reflections for $Mg_{1-x}Al_xB_2$ for selected compositions in the vicinity of the phase separation.

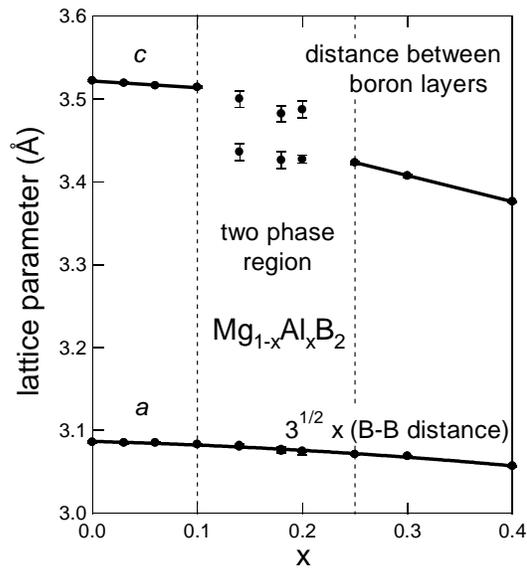

**Fig. 2**. Variation of the in-pane ($a$) and between plane ($c$) lattice parameters as a function of aluminum concentration in $Mg_{1-x}Al_xB_2$. In the two-phase region, $c$ axis values for both phases are shown.

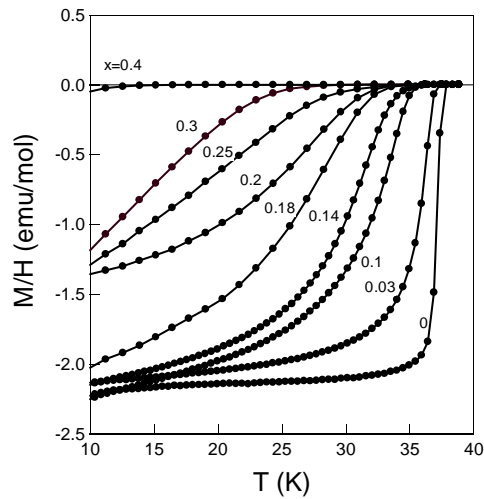

**Fig. 3**. Magnetic characterization of the superconducting transitions for polycrystalline powders of single phase $Mg_{1-x}Al_x B_2$ for x=0 to 0.40. Applied field 15 Oe, zero field cooled data. Inset: detail of the region in the vicinity of $T_c$.